\documentclass[doublecol]{epl2}

\usepackage{graphicx}
\usepackage{amssymb}
\usepackage{amsmath}
\usepackage{color}
\usepackage{comment}

\DeclareMathOperator*{\Simiq}{\simeq}

\newcommand{\ve}[1]{{\mathbf #1}}

\newcommand{\bra}[1]{\langle\left.{#1}\right|}
\newcommand{\ket}[1]{\left|{#1}\right.\rangle}

\title{Supersolidity in electron-hole bilayers with a large density imbalance}

\author{Meera M. Parish\inst{1} \and Francesca M. Marchetti\inst{2}
  \and Peter B. Littlewood\inst{1}}
\shortauthor{M. M. Parish \etal}

\institute{                    
  \inst{1} Cavendish Laboratory - JJ Thomson Avenue, Cambridge,
 CB3 0HE, United Kingdom\\
  \inst{2} Universidad Aut\'onoma de Madrid - Madrid 28049, Spain}


\abstract{We consider an electron-hole bilayer in the limit of extreme
  density imbalance, where a single particle in one
  layer interacts attractively with a Fermi liquid in the other
  parallel layer. Using an appropriate variational wave function for
  the dressed exciton, we provide strong evidence for the existence of
  the Fulde-Ferrell-Larkin-Ovchinnikov (FFLO) phase in electron-hole
  bilayers with a large density imbalance. Furthermore, within this
  unusual limit of FFLO, we find that a dilute gas of minority
  particles forms excitons that condense into a two-dimensional
  ``supersolid''.}

\begin{document}

\maketitle

\section{Introduction}
Pairing phenomena in two-component Fermi systems are a topic of
fundamental interest, having relevance to a range of fields spanning
superconductivity to QCD. For an attractive interspecies interaction,
one can have Bardeen-Cooper-Schrieffer (BCS) pairing in the
weak-coupling limit or a Bose-Einstein condensate (BEC) of
tightly-bound pairs in the strong-coupling limit.

Of particular interest is the case where the densities of the two
fermionic species are \emph{imbalanced}, so that the interspecies
pairing is then frustrated. Here, one expects more exotic pairing
scenarios such as the Fulde-Ferrell-Larkin-Ovchinnikov (FFLO)
spatially-modulated phase~\cite{Fulde1964,larkin1965}, where fermions
pair at finite centre-of-mass momentum, and both gauge invariance and
translational invariance are spontaneously broken.  However, an
unambiguous observation of the FFLO phase has remained elusive,
despite being predicted more than four decades ago.
The creation of spin-imbalanced Fermi gases in ultracold atomic
gases~\cite{zwierlein2006,partridge2006} has recently revived the hope
of realising the FFLO state, but, thus far, the atomic system has been
dominated by phase separation between superfluid and normal phases,
with FFLO only occupying a tiny sliver of the predicted phase diagram
in the three-dimensional (3D) case~\cite{sheehy2007,parish2007}.

On the other hand, electron-hole bilayers, where electrons and holes
in a semiconductor are spatially separated into two closely-spaced
quantum wells, may provide a better route to achieving the FFLO
state. Here, electrons and holes can pair to form excitons which can
then in principle condense~\cite{keldysh65,comte82}.
Such bilayers have already been successfully produced by optically
pumping coupled GaAs quantum wells (for recent experiments,
see~\cite{butov2010,snoke2009}).  More recently, independently
contacted layers have been fabricated in
GaAs~\cite{Croxall2008,Seamons2009}, where electron and hole layers
can be separately loaded by biasing and doping, thus allowing the
densities in each layer to be controlled individually and providing a
means for generating a density imbalance. The additional advantage of
considering these structures over optically-pumped coupled quantum
wells lies in the extremely long lifetimes, where the exciton decay
rate due to tunnelling recombination is essentially negligible and an
equilibrium phase diagram can be regarded as accurate.

The reason that electron-hole bilayers provide the ideal conditions
for realising the FFLO state are twofold. Firstly, the reduced
dimensionality of the bilayer system favours the FFLO phase
over the normal phase owing to
the enhanced Fermi-surface ``nesting''~\cite{parish_Q1D}, and
secondly, the intra-layer Coulomb repulsion acts to suppress any
macroscopic phase separation that might compete with the FFLO phase,
in contrast to the cold-atom system where phase separation dominates.
However, recent theoretical work on density-imbalanced electron-hole
bilayers~\cite{Pieri2007,yamashita2009,subasi2010,bishop2010}
currently paints a rather uncertain picture of FFLO. While some
studies do predict regions of FFLO in the phase diagram, they either
rely on an artificial tight-binding Hamiltonian with contact
interaction~\cite{bishop2010} or on approximations that
neglect screening~\cite{yamashita2009}. Indeed, as we will show, screening crucially
affects the stability of the FFLO phase. Other
studies~\cite{Pieri2007,subasi2010} neglect the finite momentum of
the excitons in the FFLO phase and find that FFLO is often
out-competed by other phases where the excess particles simply
coexist with the excitonic superfluid. Moreover, none of them
consider competing phases such as charge-density wave (CDW) or
Wigner crystal phases, which have been shown to be dramatically
enhanced in mass-asymmetric electron-hole
bilayers~\cite{Wigner-CDW_mass}. Clearly, further work is needed to
ascertain the existence of FFLO.

In this letter, we provide strong evidence for the existence of the
FFLO phase in an electron-hole bilayer with a large density
imbalance. To address the problem in a more controlled manner, we
consider the limit of extreme imbalance where we essentially have a
single particle in one layer interacting attractively with a Fermi
liquid in the other parallel layer. This allows us to rule out CDW
or Wigner crystal phases induced by the presence of the other layer.
It also enables us to include interactions between excitons that go beyond mean-field theory.
From this analysis, we expose an unusual \emph{bosonic} limit of FFLO,
where a dilute gas of excitons forms a condensate with a 2D spatial
modulation, a phase otherwise known as a supersolid.  Our
strong-coupling version of FFLO should also be of interest to the
communities working on strongly-correlated superconductors, where one
often encounters spatial textures coexisting with superconductivity.

\section{Model}
The basic Hamiltonian for a 2D electron-hole bilayer, $H = H_0 +
H_{int}$, consists of a kinetic part, $ H_0 = \sum_{\ve{k}\sigma}
\epsilon_{\ve{k}\sigma} c^\dag_{\ve{k}\sigma} c^{}_{\ve{k\sigma}}$,
and an interaction part ($\Omega$ is the system area):
%
\begin{multline}\label{eq:ham}
  H_{int} = \frac{1}{\Omega} \sum_{\ve{k}\ve{k'}\ve{q}} g_\ve{q}
  c^\dag_{\ve{k},1} c^\dag_{\ve{k'},2} c^{}_{\ve{k'}+\ve{q},2}
  c^{}_{\ve{k}-\ve{q},1} \\
  + \frac{1}{2\Omega} \sum_{\ve{k}\ve{k'}\ve{q}\sigma} U_\ve{q}
  c^\dag_{\ve{k}\sigma} c^\dag_{\ve{k'}\sigma}
  c^{}_{\ve{k'}+\ve{q}\sigma} c^{}_{\ve{k}-\ve{q}\sigma}\; .
\end{multline}
%
Here, the labels $\sigma = \{1,2\}$ denote the different 2D layers,
and we approximate the dispersions as quadratic,
$\epsilon_{\ve{k}\sigma} = \hbar^2\ve{k}^2/2m_\sigma$, which is
reasonable for sufficiently small momenta in GaAs quantum wells.
The bare Coulomb inter- and intra-layer interactions are
respectively given by 
\begin{align}
  U_{\ve{q}} &= \frac{2 \pi e^2}{\varepsilon q} & g_\ve{q} &= -
  U_{\ve{q}} e^{-qd}\; ,
\end{align}
where $d$ is the bilayer distance, $e$ is the electron charge, and
$\varepsilon$ is the material dielectric constant. By introducing
twice the reduced mass $m = 2(1/m_1 + 1/m_2)^{-1}$, we can define the
exciton Bohr radius $a_0 = \varepsilon \hbar^2/me^2$ and the exciton
Rydberg $E_0 =e^2/\varepsilon a_0$.
We also ignore the spin degrees of freedom and assume the layers are
completely spin-polarised by a parallel magnetic field,
but we shall return to this point later.

We focus on the problem of a \emph{single} particle in the 2nd
($\sigma=2$) layer interacting attractively with a Fermi liquid in the
1st ($\sigma=1$) layer. This minority particle can either be a hole
interacting with the electron layer or an electron interacting with
the hole layer --- the two cases are simply obtained by inverting the
mass ratio $\alpha \equiv m_2/m_1$. For a fixed bilayer distance
$d/a_0$, the only other relevant parameter is the dimensionless
  density 
$r_s \equiv 2/k_F a_0$, where $k_F=2\sqrt{\pi n_1}$ is the Fermi wave
vector of the filled 1st layer with density $n_1$.
Note that this work focuses on establishing the \emph{equilibrium}
phase diagram of a fully imbalanced electron-hole bilayer, a problem
somewhat different from the phenomenology of the x-ray-edge
singularity~\cite{Mahan}, where a dynamical transition is caused by a
sudden local perturbation, e.g.\ the emission or absorption of a
photon.  Such a dynamical transition would also require a term in the
Hamiltonian that transfers electrons between layers, which is clearly
absent in \eqref{eq:ham}.

In order to determine the phase diagram of the fully imbalanced
electron-hole bilayer, we need to establish the system ground
state. To this end, we consider the following variational state for an
excitonic quasi-particle (or electron-hole pair):
\begin{equation}
  \ket{\Psi(\ve{Q})} = \sum_{\ve{k}>k_F} \varphi_{\ve{k}\ve{Q}}
  \ c^\dag_{\ve{Q} - \ve{k},2} c^\dag_{\ve{k},1} \ket{FS}\; ,
\label{eq:molecule}
\end{equation}
where $\ket{FS}$ represents the
Fermi sea of 1-particles filled up to wave vector $k_{F}$, and we use
the notation $\sum_{\ve{k}>k_F} \equiv \sum_{\ve{k}}^{k>k_F}$.
According to its definition, the excitonic wave function
$\varphi_{\ve{k}\ve{Q}}$ has relative momentum $\ve{k}$ and
centre-of-mass momentum $\ve{Q}$.
The spread in relative momentum $\ve{k}$ of $\varphi_{\ve{k}\ve{Q}}$ is set by the inverse of the size of the exciton bound state.
Note that Eq.~\eqref{eq:molecule} gives a good description of
\emph{both} low and high density limits: In the low density limit, the
state~\eqref{eq:molecule} coincides with the exact two-body (molecular
excitonic) state in the vacuum limit ($r_s \to \infty$), while in the
high density limit, where the particles barely interact, it gives the
non-interacting ``unbound'' state:
\begin{equation} 
  \ket{\Psi_0} = c^\dag_{0,2} c^\dag_{k_F\hat{\ve{k}}, 1} \ket{FS} 
\end{equation}
corresponding to $\varphi_{\ve{k}\ve{Q}} = \delta_{\ve{k}, k_F\hat{\ve{k}}}
\delta_{\ve{Q}, k_F \hat{\ve{k}}}$. In this case, the minority particle is not bound to a majority particle and instead occupies a scattering state with well-defined (zero) momentum.
We handle the regime between these two limits by considering screened
interactions within the Random Phase Approximation (RPA).  For the
single-minority-particle case, the screened interactions
$U^{sc}_{\ve{q}}$, $g^{sc}_{\ve{q}}$ have the simple expressions
\begin{align}
  U^{sc}_{\ve{q}} &= \frac{U_\ve{q}} {1 - U_\ve{q}\Pi_{1}(\ve{q})} &
  g^{sc}_{\ve{q}} &= -U^{sc}_{\ve{q}} e^{-qd} \; ,
\end{align}
where the static polarisation operator $\Pi_1$ for the 1st layer is
given by the Lindhard function:
\begin{equation}\label{eq:RPA}
  \Pi_{1}(\ve{q}) = \frac{N_s m_1}{2\pi\hbar^2} \left[\frac{\sqrt{q^2 -
        4k_F^2}}{q} \theta (q-2k_F)-1\right] \; ,
\end{equation}
with the number of particle flavours $N_s=1$ for the spin-polarised
case. Considering screened interactions within RPA corresponds to
effectively ``dressing'' the particles in our wave function with
density fluctuations (i.e., an infinite number of particle-hole
pairs). In addition, RPA provides a good approximation in the
long-wavelength limit, $q \ll 2 k_F$, and therefore gives a good
estimate of the unbinding transition: Here, as we will show later, the
wave function $\varphi_{\ve{k}\ve{Q}}$ in~\eqref{eq:molecule} is
strongly peaked at $k=k_F$ and so the momentum transfer is small.
More generally, RPA should be reliable for sufficiently large $d$
for the pairing coupling $g^{sc}_{\ve{q}}$, since this removes
the short-range interactions between electrons and holes where RPA has
problems~\cite{Zhu1996}.

The exciton $\ket{\Psi(\ve{Q})}$ and unbound $\ket{\Psi_0}$ states
resemble, respectively, the molecule and polaron wave functions used
in ultracold atomic Fermi gases~\cite{parish2011}. But there is an
important difference: In cold atoms, one can accurately model the
phase diagram using a wave function ansatz that explicitly includes
just one particle-hole excitation on top of the non-interacting
  approximation to the Fermi sea $\ket{FS}$ for both the molecule and
polaron wave functions~\cite{combescot2009,parish2011}.  This simple
treatment is possible for cold atoms because the majority Fermi sea is
non-interacting and the interspecies interaction is short-range.
However, for long-range interactions and an interacting majority Fermi
sea, this perturbative expansion is bound to fail, as it is never
profitable to excite just \emph{one} particle-hole pair. Indeed, the
singular nature of the Coulomb interaction will generate an infinite
number of particle-hole excitations at the Fermi surface.  Thus,
instead of explicitly including a small number of particle-hole
excitations in the wave functions, we implicitly include an infinite
number of particle-hole excitations by replacing the Coulomb
potentials in Eq.~\eqref{eq:eigenvalue} with screened potentials 
and using, implicitly, an interacting Fermi sea.

\section{Phase diagram}
By minimising the expectation value $\bra{\Psi(\ve{Q})} (H-E)
\ket{\Psi(\ve{Q})}$ with respect to the amplitude
$\varphi_{\ve{k}\ve{Q}}$, we obtain an eigenvalue equation for the
exciton energy $E$:
\begin{multline}
  E \varphi_{\ve{k}\ve{Q}} = \left( \epsilon_{\ve{Q}-\ve{k},2} +
  \epsilon_{\ve{k},1} - \frac{1}{\Omega} \sum_{\ve{k}'<k_F}
  U^{sc}_{\ve{k}-\ve{k'}} \right) \varphi_{\ve{k}\ve{Q}} \\
  + \frac{1}{\Omega} \sum_{\ve{k'}>k_F} g^{sc}_{\ve{k}-\ve{k'}}
  \varphi_{\ve{k'}\ve{Q}}
\label{eq:eigenvalue}
\end{multline}
Similarly to what was done in ultracold polarised Fermi
gases~\cite{combescot2009,parish2011}, we compare the energy of the
excitonic molecular wave function $\ket{\Psi(\ve{Q})}$ with that of
the unbound state $\ket{\Psi_0}$, i.e., the exciton is bound if the
lowest eigenvalue of~\eqref{eq:eigenvalue} is such that $E < \hbar^2
k_F^2/2 m_1 - \frac{1}{\Omega} \sum_{\ve{k}'<k_F} U_{k_F \hat{\ve{k}}
  -\ve{k'}}$.

It is useful to note that the eigenvalue
equation~\eqref{eq:eigenvalue}
coincides, in the limit of full imbalance, with the mean-field gap
equation employed to describe the BEC-BCS crossover in imbalanced
electron-hole bilayers (see, e.g.,
Refs.~\cite{Pieri2007,yamashita2009,subasi2010}). To see this, we
neglect the intra-layer Coulomb repulsion and rewrite
Eq.~\eqref{eq:eigenvalue} in terms of the ``gap'',
$\Delta_{\ve{k}\ve{Q}} \equiv \frac{1}{\Omega} \sum_{\ve{k'}>k_F}
g^{sc}_{\ve{k}-\ve{k'}} \varphi_{\ve{k'}\ve{Q}}$, therefore obtaining
\begin{equation}
  \Delta_{\ve{k}\ve{Q}} = \frac{1}{\Omega} \sum_{\ve{k}'>k_F}
  g^{sc}_{\ve{k}-\ve{k'}}
  \frac{\Delta_{\ve{k}'\ve{Q}}}{E-\epsilon_{\ve{Q}-\ve{k},2} -
    \epsilon_{\ve{k},1}} \; .
\label{eq:gapeq}
\end{equation}
This corresponds to the linearised version of the mean-field gap
equation of Refs.~\cite{Pieri2007,yamashita2009,subasi2010}, which is
just the form one would expect in this limit since the gap
$\Delta_{\ve{k}\ve{Q}}$ becomes macroscropically small as we approach
full imbalance. To complete the correspondence between
Eq.~\eqref{eq:gapeq} and the linearised mean-field gap equation, we
require that the chemical potential of the minority particles be
$\mu_2 = E-\hbar^2 k_F^2/2m_1$, since the chemical potential of the
majority particles is fixed to $\mu_1 = \hbar^2 k_F^2/2m_1$.
Thus, the conditions for a bound state outlined above require that
$\mu_2 <0$, implying that $\mu_2$ gives the exciton binding energy in
this limit, which also matches with the mean-field theory.

Here, as for the excitonic wave function $\varphi_{\ve{k}\ve{Q}}$, the
order parameter $\Delta_{\ve{k}\ve{Q}}$ describes a pair with
centre-of-mass momentum $\ve{Q}$.
Therefore, in this context, it is natural to identify an exciton with
minimum energy at \emph{non-zero} momentum $\ve{Q}$ with the FFLO
phase in the large-imbalance limit --- this will be justified further
later when we consider the case of a dilute gas of minority
particles.
We emphasise that in this work, similarly to
Ref.~\cite{yamashita2009}, and contrary to
Refs.~\cite{Pieri2007,subasi2010}, where only the limit $\ve{Q} \to 0$
has been considered, we allow the pair centre-of-mass momentum to be
finite and minimise the energy with respect to $\ve{Q}$.

\begin{figure}
\begin{center}
\includegraphics[width=0.75\linewidth,angle=0]{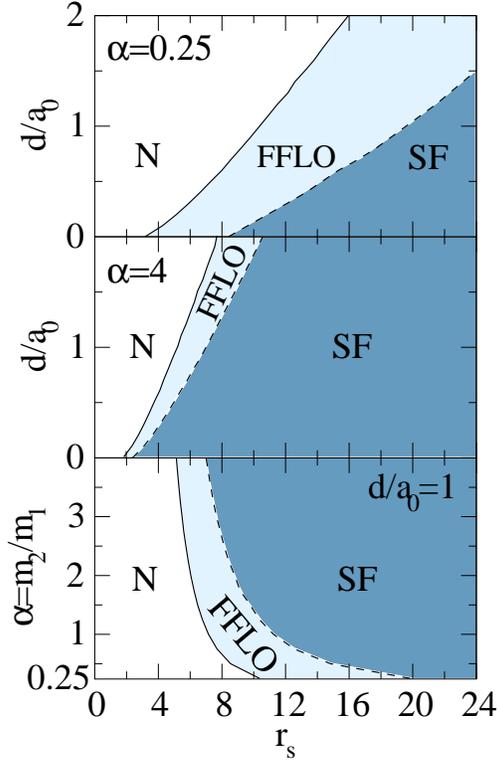}
\end{center}
\caption{
Single-minority-particle phase diagrams as a
  function of the interaction parameter $r_s$. In the top two panels,
  the bilayer distance $d/a_0$ varies and the mass ratio $\alpha =
  m_2/m_1$ is fixed to typical values in GaAs ($\alpha=0.25, 4$),
  while the bottom panel has $\alpha$ varying and $d/a_0=1$. 
  In all cases, the inter- and intra-layer interactions have been
  screened using RPA. The superfluid (SF) region corresponds to
  excitons with centre-of-mass momentum $Q=0$, while the``FFLO''
  excitons have their lowest energy when $Q\neq 0$. The normal (N)
  region is where there are no bound excitons. Refer to
  Fig~\ref{fig:Qmin} for the behaviour of $Q$ within the FFLO region
  for $d/a_0 = 1$. 
}
 \label{fig:d_vs_rs}
\end{figure}

\begin{figure}
\begin{center}
\includegraphics[width=0.75\linewidth,angle=0]{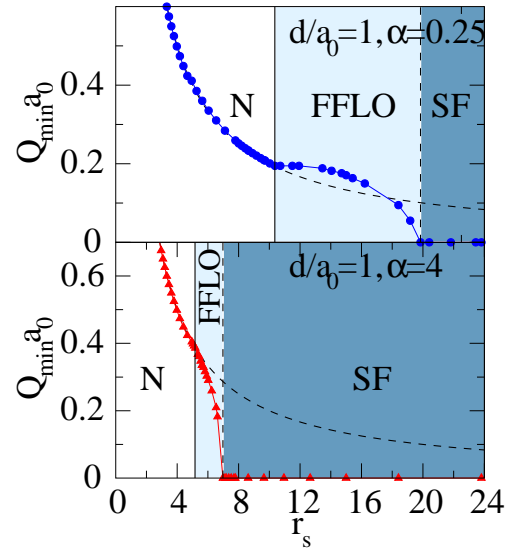}
\end{center}
\caption{
Behaviour of the momentum $Q_{min}$ that
  minimises the exciton's energy as a function of $r_s$ for fixed $d$
  and $\alpha = m_2/m_1$.  The dashed line corresponds to $Q=k_F \equiv
  2/r_s a_0$. Within the FFLO phase, $Q_{min}$ evolves continuously from
  $Q=0$ to $Q=k_F$ with decreasing $r_s$. In the normal phase, $Q_{min} =
  k_F$ because the unbound state corresponds to
  $\varphi_{\ve{k}\ve{Q}} = \delta_{\ve{k}, k_F\hat{\ve{k}}}
  \delta_{\ve{Q}, k_F \hat{\ve{k}}}$.}
 \label{fig:Qmin}
\end{figure}

The phase diagram for the fully-imbalanced bilayer system is obtained
by converting the eigenvalue equation~\eqref{eq:eigenvalue} into a
matrix equation and numerically solving for the lowest energy
eigenvalue. We fix the electron-hole mass ratio to $4$, which is
approximately its value in GaAs experiments, so that the relevant mass
ratios are $\alpha = \{0.25, 4\}$.
In the low-density limit, $r_s \rightarrow \infty$, we recover the
two-body limit, and thus we expect a bound exciton with $Q=0$ (which
we refer to as SF) for sufficiently large $r_s$, as shown in
Fig.~\ref{fig:d_vs_rs}.
In the opposite limit, where $r_s$ is small, we see that the the
screened interactions cause the exciton to eventually unbind and enter
the ``normal'' (N) phase, as expected. However, the crucial point is
that in a significant region of the phase diagram at intermediate
densities, the system ground state is a bound exciton with finite
momentum $Q$, which we label as FFLO, as previously explained.  The
size of the FFLO region is greatest if the minority particle is an
electron ($\alpha = 0.25$) rather than a hole ($\alpha = 4$).  Indeed,
the FFLO region is generally enhanced (and shifted to larger $r_s$)
when the minority particle is lighter, as demonstrated in the bottom
panel of Fig.~\ref{fig:d_vs_rs}.
The reason for this is simple: an exciton
with $Q=0$ requires the minority particle to sit above the Fermi
sea, but a small $\alpha$ 
increases the kinetic energy
cost for this, thus favouring the formation of an FFLO exciton, where
the minority particle can sit below the Fermi surface.

The FFLO region is also enlarged by increasing the bilayer distance $d$ (Fig.~\ref{fig:d_vs_rs}). 
For large $d$, large momentum
scattering $|\ve{k} - \ve{k}'| > 1/d$ is suppressed in
$g^{sc}_{\ve{k} - \ve{k}'}$ and this tends to favour FFLO, where the
wave function $\varphi_{\ve{k}\ve{Q}}$ is peaked in the direction of
$\ve{Q}$, over SF. However, larger $d$ also demands larger $r_s$ to
achieve FFLO, and so we eventually expect Wigner crystallisation in
the 1st layer to destroy FFLO.
According to estimates from Quantum Monte Carlo (QMC) calculations~\cite{ceperley1989,bachelet2002}, Wigner crystallization occurs once $r_s \gtrsim 70\alpha/(1+\alpha)$. Thus, the distance $d$ required to see FFLO sensitively depends on $\alpha$, 
e.g.\ for $\alpha=0.25$, we ideally want $d/a_0 \lesssim 1$.

Finally, we note that both the SF-FFLO and FFLO-N transitions are
second order, with the exciton momentum $Q$ varying continuously
(Fig.~\ref{fig:Qmin}). Also, we always find that $Q=k_F$ at the
FFLO-N transition.

Interestingly enough, we find that the size of the FFLO region
actually strongly depends on the screening of the interaction. In
order to understand this, we consider the gap-equation form of the
eigenvalue equation~\eqref{eq:gapeq} for the case of unscreened
interactions.  After converting sums into integrals, and rescaling
momenta by $a_0^{-1}$ and energies by $E_0$, we obtain:
\begin{equation}
  \Delta_{\ve{k}\ve{Q}} = \int_{k'>\frac{2}{r_s}}
  \frac{d^2\ve{k'}}{2\pi} \frac{e^{-d
      |\ve{k}-\ve{k'}|}}{|\ve{k}-\ve{k'}|}
  \frac{\Delta_{\ve{k'}\ve{Q}}}{-E + \epsilon_{\ve{Q}-\ve{k'},2} +
    \epsilon_{\ve{k'},1}}
\label{eq:unscreen}
\end{equation}
Now, at the unbinding transition [$E =4\alpha/r_s^2(1+\alpha)$ in
  rescaled units], we see that that the integral is logarithmically
divergent for $\ve{Q} = \ve{k'} = \ve{k} = 2\hat{\ve{k}}/r_s$ and so we
must take $r_s = 0$ for the equation to be satisfied. This implies
that, for the bare Coulomb inter-layer interaction $g_{\ve{q}}$, the
exciton with momentum $Q = 2/r_s \equiv k_F$ for $r_s \ll 1$ is
\emph{always} bound, a point which we have also confirmed
numerically. In other words, FFLO will extend all the way down to $r_s
= 0$ in the phase diagram, which is contrary to what was found
in~\cite{yamashita2009}. Note that this is \emph{not} the case when
the interaction is screened, since this removes the singularity at
$\ve{k'} = \ve{k}$, thus leaving an integrable singularity at $\ve{Q}
= \ve{k'} = 2\hat{\ve{k}}/r_s$.

\section{Dilute gas of minority particles}
To understand the implications of our single-minority-particle phase
diagram for a dilute gas of minority particles (and, thus, the
density-imbalanced electron-hole bilayer), we must determine the
interaction between minority particles. Firstly, in the normal phase,
the effective interaction $V^{22}_{\ve{q}}$ between two unbound
minority particles within RPA satisfies $V^{22}_{\ve{q}} = U_\ve{q} +
g_{\ve{q}} \Pi_1 g^{sc}_{\ve{q}}$~\cite{zheng94} and therefore is
given by
\begin{equation}
  V^{22}_{\ve{q}} = U_\ve{q} + \frac{g^2_{\ve{q}} \Pi_1}{1 - \Pi_1
    U_\ve{q}}\; .
\end{equation}
In the long-wavelength limit ($q \rightarrow 0$), this interaction is
repulsive and dipolar ($V^{22}(r) \sim 1/r^3$), and thus we expect a
dilute gas of minority particles (with density $n_2 \ll 1/d^2$
when $d \gg 1$) to form a weakly-interacting Fermi liquid.
This is similar to what was found in Ref.~\cite{aleiner_fullpol} for
the minority spins in a strongly-polarised 2D electron gas.

The interaction between well-separated excitons is also easily
calculated by treating the excitons as static dipoles and summing up
the inter- and intra-layer contributions:
\begin{align}
  \displaystyle V^{ex}_{\ve{q}} & = 2 g^{sc}_{\ve{q}} +
  U^{sc}_{\ve{q}} + V^{22}_{\ve{q}} \Simiq_{q \to 0} 4 \pi d (1 - qd)
\label{eq:exc_int}
\end{align}
The exciton-exciton interaction in the vacuum is clearly repulsive and
dipolar at large distances, and we see from Eq.~\eqref{eq:exc_int}
that this remains the case in the presence of a Fermi sea. This
suggests that FFLO excitons in the regime of large density-imbalance
will be thermodynamically stable against phase separation, unlike in
the cold-atom case. However, note that our approximation for
$V^{ex}_{\ve{q}}$ is no longer valid when the exciton size becomes
comparable to the distance between minority particles, which is the
case near the FFLO-N transition. There is also the possibility that
the short-range exciton-exciton interactions are attractive, in which
case there will be biexciton formation. However, QMC
calculations for the two-exciton problem show that biexcitons cannot
form when $d \gtrsim 0.25$ for mass ratio $\alpha =
4$~\cite{biexciton}. Indeed, biexciton formation will be even more
suppressed in our spin-polarised case due to Pauli exclusion.

Now that we have an estimate for the exciton-exciton interaction, we
can determine the structure of the FFLO phase at large density
imbalance. For a low density of minority particles, we can treat the
excitons as simple bosons and define the mean-field complex order
parameter $\psi(\ve{r})$, where $|\psi(\ve{r})|^2$ corresponds to the
exciton density. Fluctuations will, of course, prevent true long-range
order at finite temperature in 2D, but our excitonic phases can still possess
quasi-long-range order.

The effective low-energy thermodynamic potential $F[\psi]$ is similar
to that used in weak crystallisation theory~\cite{weak_crys}:
\begin{equation*}
  F[\psi] = \int d^2\ve{r} \left[-\mu |\psi|^2 + \gamma
  \left|(\nabla^2+Q_{min}^2)\psi \right|^2 +\frac{\lambda}{2}
  |\psi|^4\right] \; .
\end{equation*}
Here, $\mu$ is the chemical potential for the excitons, where $\mu <
0$ describes an empty 2nd layer, while the term $\gamma > 0$ sets the
exciton's dispersion minimum at momentum $Q_{min}$. 
The repulsive
interactions are taken to be contact, with $\lambda =
V^{ex}_0$ 
--- the momentum dependence of $V^{ex}_{\ve{q}}$ should not affect 
our solution below provided $Q_{min} d \ll 1$, i.e.\ the typical length scale of the spatial modulations is large. Indeed, this condition is always satisfied within the FFLO regions of Fig.~\ref{fig:d_vs_rs}.
For the FFLO phase, we consider solutions of the form 
%
  $\psi(\ve{r}) = \sum_n a_n e^{i \ve{q}_n \cdot\ve{r}}$, 
with $|\ve{q}_n| = Q_{min}$. Substituting this into $F[\psi]$ 
we get
\begin{multline*}
  \frac{F[\psi]}{\Omega} = -\mu A +
  \frac{\lambda}{2}\left(2A^2\right.\\
  \left. -\sum_n |a_n|^4 + B^*B-\sum_n|a_n|^2|a_{-n}|^2\right) \; ,
\end{multline*}
where $A=\sum_n |a_n|^2$ and $B=\sum_n a_n a_{-n}$. Minimising with
respect to the amplitude $a_n$ we find that the lowest energy solution
($F[\psi_0]/\Omega = -\mu^2/3\lambda$) requires that $B^* a_1= |B|
a_{-1}^*$, $B^* a_2= -|B| a_{-2}^*$,
$|a_1|=|a_2|=\sqrt{\mu/6\lambda}$, while $a_n=0$ for $n>2$. Such a
state is thus composed of four $\ve{q}_n$, i.e. $\pm \ve{q}_1$ and
$\pm \ve{q}_2$, so that we have:
\begin{equation}
  \psi_0(\ve{r}) = e^{i\theta} \sqrt{\frac{2\mu}{3 \lambda}}
  \left[\cos(\ve{q}_1 \cdot \ve{r}) - i\cos(\ve{q}_2 \cdot \ve{r} +
    \theta_{12})\right] \; ,
\label{eq:quadr}
\end{equation}
 Note
that the phases $\theta$, $\theta_{12}$ and the directions
$\hat{\ve{q}}_1$, $\hat{\ve{q}}_2$ are randomly chosen.
Equation~\eqref{eq:quadr} corresponds to an exciton condensate with
a 2D spatial modulation --- a supersolid.
Contrast this with the opposite limit of weak polarization in the
BCS regime, where the favored state is believed to be a
\emph{single} cosine~\cite{larkin1965}.

\section{Discussion}
The supersolid is expected to enjoy a sizeable region of existence
away from Wigner crystallisation, but one possible issue in the
large imbalance limit is the formation of three-body states, or
trions. These are known to exist for all mass ratios in the limits
$d \to 0$ and $k_F \to 0$~\cite{Stebe1989}. However, we expect
trions to disappear with increasing $d$, like in the biexciton case,
since the intra-layer repulsion will eventually dominate over the
inter-layer attraction. Moreover, we find that the spin-polarised
trion is barely bound at $d = 0$ and so we expect trions to exist
only for $d \ll 1$ in Fig.~\ref{fig:d_vs_rs}. We speculate that a
spin-polarised electron-hole bilayer (as considered in this work)
may be better for achieving the FFLO phase than an unpolarised one,
since spin polarisation both suppresses trion formation and enhances
the inter-layer attraction, due to the reduction in screening in
Eq.~\eqref{eq:RPA}.

Thus far, our calculations have been restricted to zero temperature and we have ignored the effects of thermal fluctuations. However, the exciton binding energy $E_B$ provides a temperature scale below which the exciton should be robust against thermal fluctuations. Assuming that the minority particles are electrons ($\alpha\simeq 0.25$) and using the parameters in
GaAs ($a_0 \simeq 7$~nm and $E_0 \simeq 17$~meV), we find that $E_B$ 
for FFLO excitons at $d/a_0 \simeq 0.5$ is of the
order of $5$~K near the SF-FFLO transition. Thus, FFLO excitons should be experimentally accessible.

In order to access the FFLO phase itself, we require a sufficiently large exciton density, i.e.\ a sufficiently large minority-particle density $n_2$, since we expect the critical temperature to scale with $n_2$. Our predictions should be valid at finite $n_2$ provided the exciton size is smaller than the spacing between excitons, i.e.\ we must have $n_2/n_1 < r_s^2 E_B/E_0$. Thus, our strong-coupling theory of FFLO should 
apply for $n_2/n_1 <1$ deep within the FFLO region, e.g.\ for $d/a_0 \sim 0.5$ and $r_s \sim 10$, but it will
break down near the FFLO-N transition for finite $n_2$. Here, we expect a more BCS-like 
version of FFLO, with the FFLO-N transition shifting to lower $r_s$ as $n_2$ is increased~\cite{yamashita2009}.

For an ordinary excitonic superfluid in 2D, there will be a Berezinskii-Kosterlitz-Thouless transition~\cite{KTtransition,berezinskii1972} to the superfluid state, with transition temperature given by\footnote{Equation~\eqref{eq:KT} assumes all the excitons are superfluid and neglects the effect of interactions, but this is only expected to reduce $T_{BKT}$ by of order 10\% in the region of interest according to QMC calculations for bosons with dipolar interactions~\cite{filinov2010}.}
\begin{align} \label{eq:KT}
  T_{BKT} = \frac{\alpha}{(1+\alpha)^2} \frac{E_0}{k_Br_s^2} \frac{n_2}{n_1}
\end{align}
It is unclear whether or not this is also true for the FFLO phase, since translational as well as gauge invariance has been spontaneously broken. However, on general grounds, we expect the FFLO transition temperature to have the same scaling as $T_{BKT}$, so we can use to it to obtain an estimate. 
For the typical parameters $r_s\sim10$ and $n_2/n_1 \sim0.2$, we get an FFLO transition temperature of order 100mK, which is smaller than the exciton binding energy (as expected), but still within reach experimentally.

The FFLO phase we predict can be observed experimentally via light
scattering off of the spatial modulations. In addition, if the electrons and holes
are allowed to recombine, then a signature of the finite momentum
pairing will appear in the angular emission of the
photons~\cite{angular_emission}, where we expect peaks in the
exciton photoluminescence to occur at large angles with respect to
the plane normal, corresponding to $\pm \ve{q_1}$ and $\pm
\ve{q_2}$.

\acknowledgments We are grateful to K. Gupta and D. Khmelnitskii for
fruitful discussions. MMP acknowledges support from the EPSRC. FMM
acknowledges financial support from the programs Ram\'on y Cajal and
Intelbiomat (ESF). This work is in part supported by the Spanish MEC
QOIT-CSD2006-00019.


\end{document}